\documentclass[prl,showpacs,twocolumn]{revtex4}
\usepackage{graphicx}
\usepackage{amsmath}
\usepackage{amsfonts}
\begin{document}
\title{Driving phase slips in a superfluid atom circuit with a rotating weak link}
\author{K. C. Wright}
\altaffiliation{Present Address: Department of Physics and Astronomy, Dartmouth College, NH, 03755, USA}
\author{R. B. Blakestad}
\altaffiliation{Present Address: Booz Allen Hamilton, Arlington, VA 22203, USA}
\author{C. J. Lobb}
\altaffiliation{Also at the Center for Nanophysics and Advanced Materials, University of Maryland, College Park, MD, 20742, USA}
\author{W. D. Phillips}
\author{G. K. Campbell}
\affiliation{Joint Quantum Institute, National Institute of Standards and Technology and University of Maryland, Gaithersburg, MD, 20899, USA}

\begin{abstract}
We have observed well-defined phase slips between quantized persistent current states around a toroidal atomic ($^{23}$Na) Bose-Einstein condensate. These phase slips are induced by a weak link (a localized region of reduced superfluid density) rotated slowly around the ring. This is analogous to the behavior of a superconducting loop with a weak link in the presence of an external magnetic field. When the weak link is rotated more rapidly, well-defined phase slips no longer occur, and vortices enter into the bulk of the condensate. A noteworthy feature of this system is the ability to dynamically vary the current-phase relation of the weak link, a feature which is difficult to implement in superconducting or superfluid helium circuits.
\end{abstract}

\maketitle

Many quantum fluids exhibit superfluid phenomena, including dissipationless flow and persistent circulation. Furthermore, a quantum fluid interrupted by a weak link (\emph{e.g.}, a Josephson junction) can act as a nonlinear interferometer, allowing the creation of highly sensitive detectors such as SQUID magnetometers~\cite{ClarkeSQUID04} and superfluid helium gyroscopes~\cite{PackardPrinciplesPRB92, SchwabDetectionN97, AvenelDetectionPRL97, SimmondsQuantumN01, HoskinsonSuperfluidPRB06, SatoSuperfluidRoPiP12}. Ultracold atomic gases offer new opportunities for control and measurement~\cite{LeggettQuantum06}, and it is now becoming possible to realize ultracold atomic ``circuits'' in configurations analogous to those mentioned above~\cite{RamanathanSuperflowPRL11}. Such devices provide opportunities for refining our understanding of flow and dissipation in quantum fluids, and may also prove useful for inertial sensing.

Several experiments with ultracold atomic gases in simply-connected geometries have observed a critical velocity for dissipation when a defect is moved through the system~\cite{RamanEvidencePRL99, OnofrioObservationPRL00, InouyeObservationPRL01, EngelsStationaryPRL07, MillerCriticalPRL07, NeelyObservationPRL10, DesbuquoisSuperfluidNP12}. Other experiments have observed Josephson effects across a thin barrier~\cite{AlbiezDirectPRL05, LevyACDCN07, LeBlancDynamicsPRL11}. More recent experiments with Bose-Einstein condensates in multiply-connected geometries have demonstrated persistent currents~\cite{RyuObservationPRL07}, the stochastic decay of persistent currents in a series of quantized steps~\cite{MoulderQuantizedPRA12}, and phase slips across a stationary weak link~\cite{RamanathanSuperflowPRL11}.

\begin{figure}
\includegraphics[width=3.2in]{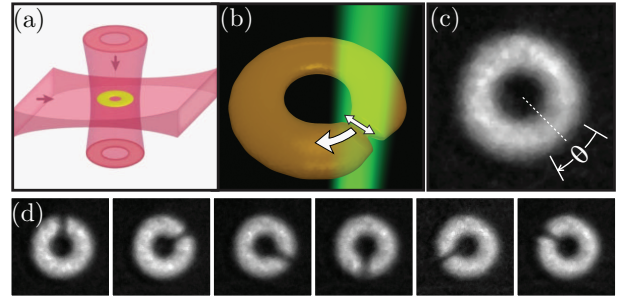}
\caption{Ring condensate and the weak link created in it by the repulsive potential of a blue-detuned ($\lambda$ = 532 nm) laser field. (a) Schematic showing the attractive optical dipole trap formed by a red-detuned ($\lambda$ = 1064 nm) horizontal ``sheet'' beam and a vertical ``ring'' beam. (b)  Geometry of the ``barrier'' beam. The small double-ended arrow indicates rapid ($2$ kHz) radial scanning of the beam. The larger single-ended arrow indicates the slow (azimuthal) rotation at up to 3 Hz.  (c) \emph{in situ} absorption image of the ring condensate, viewed from above. The arc of the barrier trajectory, $\theta$, was centered over the azimuth of lowest density, indicated by the dotted line. The field of view is 84 $\mu$m square. (d) Sequence of \emph{in situ}  images showing the effect of the barrier at successive 60$^{\circ}$ intervals around the ring. The barrier height is $\approx$ 60 \% of the chemical potential. Each image in (d) is the average of three absorption images, with a 93 $\mu$m square field of view.}
\label{ringwithbarrier}
\end{figure}

In this Letter, we demonstrate \emph{deterministic} phase slips between quantized circulation states, caused by \emph{rotating} a weak link around a toroidal (\emph{i.e.} annular) condensate (see Fig.~\ref{ringwithbarrier}). In the rotating frame, the response of the system is analogous to that of a static superconducting ring, with a Josephson junction, in an external magnetic field. In this analogy, the quantized circulation of the superfluid corresponds to magnetic flux, and the rotation of the barrier corresponds to an external magnetic field. Rotation of the barrier causes a supercurrent $I$ to flow through it. This flow is dissipationless below some critical current $I_\mathrm{c}$, leaving the quantized circulation state in the ring unchanged. At a high enough rotation rate $I$ exceeds $I_\mathrm{c}$, and excitations can occur. For some circuit parameters, these excitations lead to a change in circulation state via a discontinuous jump in the phase of the condensate wavefunction, \emph{i.e.}, a phase slip~\cite{AvenelObservationPRL85, AvenelJosephsonPRL88, AmarQuantizedPRL92}.  Increasing the rotation rate again until $I$ exceeds $I_\mathrm{c}$ will cause another phase slip. The sharp change of circulation (or magnetic flux) in the ring in response to rotation (or magnetic field) is the key to the exceptional sensitivity of devices such as SQUID magnetometers~\cite{ClarkeSQUID04} and superfluid helium gyroscopes~\cite{SatoSuperfluidRoPiP12}.

The superfluid state of the condensate is given by $\psi(\vec{r}) = \sqrt{\rho(\vec{r})}e^{i\phi(\vec{r})}$, where $\rho$ is the superfluid density, and $\phi$ is the phase. In an inertial frame $\phi$ is related to the superfluid velocity by $\vec{v}_\mathrm{s} = \frac{h}{2\pi m}\nabla\phi$, where $h$ is Planck's constant and $m$ is the atomic mass. The circulation around any closed path, $P$, must be quantized according to the Bohr-Sommerfeld rule, $\oint_P \vec{v}_\mathrm{s}\cdot d\vec{l} = n \kappa_0 $, where $n$ is the integer winding number of the circulation, and $\kappa_0 = h/m$. In a ring geometry without a barrier, this gives rise to a series of metastable persistent current states, with the ground state having $n=0$.

When two bulk coherent regions are coupled by a tunnel junction, the current is a $2\pi$ periodic function of the phase difference $\gamma$ across the junction~\cite{ClarkeSQUID04}. Even non-tunnel coupling can lead to single-valued and $2\pi$ periodic current-phase relations under certain circumstances~\cite{LikharevSuperconductingRMP79}. The weak link in our experiments is not a tunnel junction; in the direction of flow it is thick  compared to the condensate healing length, and is more analogous to a superconducting constriction (\emph{e.g.}, Dayem bridge)~\cite{AndersonRadio-FrequencyPRL64}. The current-phase relation in such a geometry is in general not single-valued or $2\pi$ periodic, but can become so when the peak height of the barrier potential, $U_\mathrm{b}$, is on the order of the condensate chemical potential, $\mu_0$~\cite{BaratoffCurrent-PhasePRL70, DeaverRelaxationPLA72, PiazzaCurrent-phasePRA10}. As discussed below, our present data are consistent with the current-phase relation of an ideal Josephson junction in series with a small linear inductance.

We create condensates of $\approx\!6\times 10^5$ $^{23}$Na atoms in a toroidal optical dipole trap~\cite{RamanathanSuperflowPRL11} (see Fig.~\ref{ringwithbarrier}a). The radius of the ring-shaped potential minimum, which coincides with the peak atomic density, is determined by absorption imaging to be $R=19.2(3)$ $\mu$m~\cite{endnote:uncertainty}. The radial Thomas-Fermi half-width of the annulus is $w=10.5(8)$ $\mu$m. We estimate the vertical half-width to be $\approx2$ $\mu$m using the measured vertical trap frequency of 600(5) Hz and a calculated chemical potential $\mu_0/h\approx2$ kHz. The azimuthal variation of the trap depth is $<5\%$ of $\mu_0$, resulting in the nearly uniform density profile shown in Fig.~\ref{ringwithbarrier}(c). The condensate is initially formed in a non-rotating state~\cite{endnote:spontaneous}.

The rotating weak link was created by a focused, blue-detuned laser field, which caused the atoms to experience a localized (optical dipole) force repelling them from the beam. The focus was a spot $\approx8$ $\mu$m in diameter (FWHM), which was moved along a chosen trajectory in the plane of the ring, using a 2-axis acousto-optic deflector. Rapid scanning of the spot allows us to create effective time-averaged optical potentials~\cite{HendersonExperimentalNJP09}. In this experiment we created a wide, flat barrier by scanning the radial position of the spot across the condensate as a triangle-wave, with a scan amplitude greater than 2$w$ (see Fig.~\ref{ringwithbarrier}b). We rotated this time-averaged barrier azimuthally for 1.5 s at constant $\Omega/2\pi\leq$ 3 Hz, which is $<1/10$ the angular frequency of sound propagating around the ring. During the first 0.5 s, the strength of the barrier potential was increased linearly to a value $U_\mathrm{b} \approx \mu_0/2$. It was then held constant for 0.5 s, and finally ramped off over the last 0.5 s. To minimize systematic effects due to variations in the effective $I_\mathrm{c}$ around the ring, we determined the region of lowest density and centered the barrier trajectory over this region (Fig.~\ref{ringwithbarrier}(c)).

The critical current through the weak link, $I_\mathrm{c}$, is reached when the superfluid velocity becomes equal to the critical velocity~\cite{RamanathanSuperflowPRL11, WatanabeCriticalPRA09}. Increasing the barrier height reduces the local superfluid density, decreasing the critical current. The ability to dynamically vary the strength of the weak link and its current-phase relation is an advantageous feature of this atom circuit, potentially allowing for ``third-terminal'' functionality that is difficult to achieve in superconducting circuits~\cite{BaselmansReversingN99, BellControllableAPL03, NevirkovetsEnhancementPRB07}.

\begin{figure}
\includegraphics[width=3.2in]{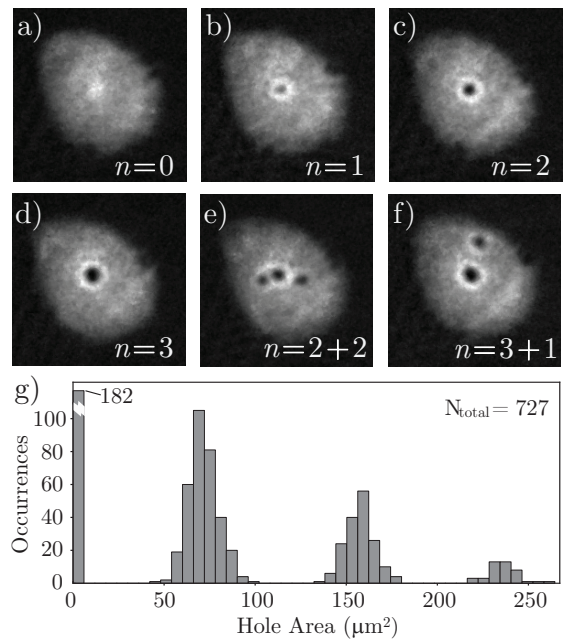}
\caption{Quantization of circulation in the ring condensate. (a-f) Absorption images of the condensate after 10 ms time-of-flight expansion, showing examples of different circulation states, $n$. The field of view is 180 $\mu$m square. The central hole is the signature of the persistent current phase singularity, which is trapped in the center of the ring prior to release. (e,f) The off-center holes are due to vortices within the annulus.  (g) Histogram of hole size distribution, showing grouping around discrete values due to quantization of circulation in the condensate.  The vertical axis is the number of occurrences per bin, the horizontal axis is the area of the hole~\cite{supplemental1}. This histogram shows the same data set used to generate Fig.~\ref{rotationstep}}
\label{histogram}
\end{figure}

To determine the circulation state of the condensate, we released it from the trap, allowing it to expand in time-of-flight (TOF). We then imaged the condensate, looking perpendicular to the plane of the ring. A condensate in a non-circulating state expands into a smooth density profile that is peaked at the center, as shown in Fig.~\ref{histogram}(a). 

Circulation in the condensate manifests itself in two different ways. A pure persistent current state exhibits azimuthal flow around the ring, with no vortex core present in the annulus. In TOF, this results in a single central hole with a size that increases with the winding number of the current around the ring, as shown in Fig.~\ref{histogram}(b-d). At high rotation rates, vortices are expected~\cite{DonnellyStabilityPRL66, FetterLow-LyingPR67} to enter the annular region and can remain there, appearing as off-center holes in the TOF density profile. These are distinguishable from the central hole caused by a persistent current, as shown in Fig.~\ref{histogram}(e,f).

\begin{figure}
\includegraphics[width=3.2in]{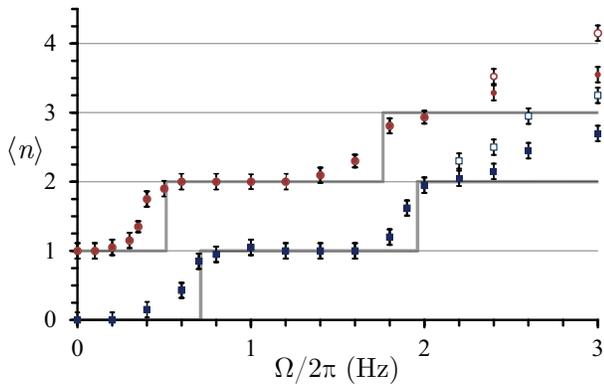}
\caption{Phase slips between quantized circulation states in the ring condensate. The vertical axis is the average winding number, $\left<n\right>$, as a function of barrier angular frequency, $\Omega$, for two barrier heights: $U_\mathrm{b}$ = 0.50 $\mu_0$ (blue squares), and $U_\mathrm{b}$ = 0.57 $\mu_0$ (red circles, vertically offset by 1 unit for clarity). The solid symbols show $\left<n\right>$, as determined from the central hole size. The open symbols are $\left<n\right>$, plus the average number of annular vortices. The number of runs averaged for each data point is at least 20; the error bars are an estimate of the 1$\sigma$ confidence interval~\cite{endnote:stepuncertainty}. The grey lines show the positions at which phase slips are expected for $\beta_L = 2$ (lower) and $1$ (vertically offset), with $\Omega_0/2\pi = 1.26$ Hz. These values of $\beta_L$ were chosen to match the $0\rightarrow 1$ steps.}
\label{rotationstep}
\end{figure}

The histogram in Fig.~\ref{histogram}(f) shows the central hole size distribution~\cite{endnote:ringrelax} for the data in Fig.~\ref{rotationstep}, comprising 727 experimental runs. The quantization of circulation is evident from the peaking of the distribution about a series of values, as has been previously noted~\cite{HendersonExperimentalNJP09, MoulderQuantizedPRA12}. The width of the peaks is due to variation in the TOF expansion dynamics,  primarily caused by shot-to-shot atom number fluctuations. Annular vortices are not stable in the absence of the rotating barrier, and leave the annulus on a time scale that depends on the condensate temperature, the trap geometry, and trap smoothness~\cite{FedichevDissipativePRA99, Abo-ShaeerFormationPRL02}. Under the conditions of our experiment, the measured lifetime of these annular vortices is about 3 seconds, therefore an annular vortex formed during our 1.5 second stirring procedure is likely to be present during TOF imaging.
 
Using the TOF images, we can determine the change in circulation caused by the barrier rotating at a given angular frequency $\Omega$. Figure~\ref{rotationstep} shows the response of the condensate to the rotating barrier for two barrier heights. The lower data set is for a $U_b=0.50(7)$ $\mu_0$, the upper data set is for a barrier height 14(2)\% higher. For each $U_\mathrm{b}$ and $\Omega$, we repeated the experiment at least 20 times, and averaged the measured winding numbers $n$. For small $\Omega$ the condensate is unperturbed, and $n=0$ after the stirring procedure. As $\Omega$ increases, the average winding number $\left<n\right>$ changes from 0 to 1. The value of $\left<n\right>$ remains 1 up to a higher angular frequency, where a transition to $\left<n\right>=2$ occurs. A comparison of the two data sets shows that by varying the barrier height $U_b$ we can adjust the critical current $I_c$. In particular, a lower $U_b$ corresponds to a higher $I_c$, \emph{i.e.} a higher critical $\Omega$. 

\begin{figure}
\includegraphics[width=3.2in]{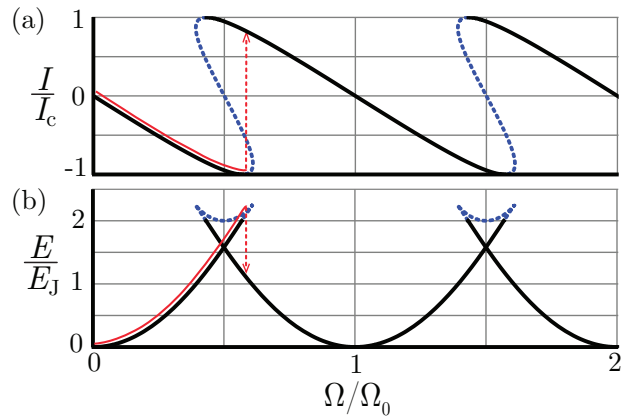}
\caption{  Normalized current $I/I_\mathrm{c}$ (a) and circulation energy $E/E_J$ (b) from equations \eqref{eq:rotquant2} and \eqref{eq:energy}, as a function of the normalized rotation speed $\Omega/\Omega_0$ ($\Omega_0 = \hbar/mR^2$), for $\beta_L = 2$. For this $\beta$, $E_J/h \approx$ 30 kHz. The solid black line is the dynamically stable region for $n=0$. The dotted lines show the unstable regions. The solid blue lines are the stable solutions for $n=\pm 1$. The red arrows indicate the expected evolution of a system initially in the $n=0$ state as $\Omega$ is increased from 0 up to a critical point $\Omega_\mathrm{c}$ at which a (dissipative) phase slip to $n=1$ occurs.}
\label{current_energy}
\end{figure}

To model our experiment, we consider a 1D system in a frame co-rotating with the weak link. In this rotating frame, the superfluid velocity is $\vec{v}_\mathrm{s}\!' = \vec{v}_\mathrm{s} - \vec{\Omega}\times\vec{r}$, and the Bohr-Sommerfeld quantization condition is
\begin{equation}
2\pi n = 2\pi\frac{\kappa(\Omega)}{\kappa_0}+ 2\pi \frac{L_R I}{\kappa_0}+\gamma(I)\label{eq:rotquant}.
\end{equation}
Here $\kappa(\Omega) = 2 \pi \Omega R^2$ is the external ``circulation flux'' due to rotation at angular velocity $\Omega$, $L_R= l/ \rho_{\mathrm{1D}}$ is the kinetic/hydrodynamic inductance of the ring, excluding the junction~\cite{KadinIntroduction99, SatoSuperfluidRoPiP12}, $l$ is the circumference of the ring minus the effective length of the junction, and $\rho_{\mathrm{1D}}$ is the mass per unit length around the ring, which is assumed constant outside the barrier region. $I$ is the mass current in the rotating frame.  The terms on the right of Eq.~\eqref{eq:rotquant} are, respectively, the Sagnac phase due to the rotating frame, the change in phase due to current around the ring, and the phase drop across the weak link. The last term is determined by the current-phase relation of the weak link, which we approximate as an ideal sinusoidal current-phase relation plus a small linear kinetic inductance $L_\mathrm{b}$~\cite{DeaverRelaxationPLA72, BaratoffCurrent-PhasePRL70, PiazzaCurrent-phasePRA10}:
\begin{equation}
\gamma(I) = \sin^{-1}\left(\frac{I}{I_\mathrm{c}}\right)+ 2 \pi \frac{L_\mathrm{b} I}{\kappa_0}.\label{eq:currentphase}
\end{equation}
We combine equations \eqref{eq:rotquant} and \eqref{eq:currentphase} and obtain
\begin{equation}
n = \frac{\kappa(\Omega)}{\kappa_0} + \frac{1}{2\pi}\left[\beta_L\frac{I}{I_\mathrm{c}} + \sin^{-1}\left(\frac{I}{I_\mathrm{c}}\right)\right] \label{eq:rotquant2}.
\end{equation}
The parameter $\beta_L = L_H/L_J$ is the ratio of the total hydrodynamic inductance $L_H = L_R + L_b$ to the (fluid) Josephson inductance of the weak link $L_J = \kappa_0/2\pi I_c$. Figure~\ref{current_energy}(a) is a normalized plot of $I(\Omega)$ as implicitly defined by Eqn.~\eqref{eq:rotquant2}, with $\beta_L = 2$, which is in the hysteretic regime~\cite{endnote:hysteresis}, and approximates the conditions of our experiment for $U_\mathrm{b}$ = 0.5 $\mu_0$. The circulation energy, $E$, can be written as~\cite{Tinkham}:
\begin{equation}
\frac{E}{E_J} = \frac{1}{2}\beta_L \left(\frac{I}{I_\mathrm{c}}\right)^2+\left[1-\cos\left( 2\pi \frac{\kappa(\Omega)}{\kappa_0}+ \beta_L \frac{I}{I_\mathrm{c}}\right)\right],  \label{eq:energy}
\end{equation}
where $E_J = \kappa_0 I_c /2 \pi$ is the Josephson energy of the weak link. Equations~\eqref{eq:rotquant2} and~\eqref{eq:energy} implicitly define $E(\Omega)$, plotted in Fig.~\ref{current_energy}(b), with $\beta_L=2$. For each value of the winding number $n$, the stable branches (solid lines)  continue up to $I=I_\mathrm{c}$.  Above this point, the system becomes dynamically unstable and dissipation can occur (dashed lines). At this point in the plots of Fig.~\ref{current_energy}, there is a state of different $n$ with lower total energy, and a phase slip occurs, taking the system to that state. The critical angular frequency $\Omega_\mathrm{c}^{\pm}$ at which a phase slip occurs to $n\pm1$ is found by setting $I/I_\mathrm{c}=\pm 1$ in Eqn.~\eqref{eq:rotquant2}, and solving for $\Omega$:
\begin{equation}
\frac{\kappa_\mathrm{c}^{\pm}}{\kappa_0}=\frac{\Omega_\mathrm{c}^{\pm}}{\Omega_0} =n\pm\left(\frac{1}{4}+\frac{\beta_L}{2\pi}\right),
\end{equation}
where $\Omega_0 = \kappa_0/2 \pi R^2$. The solid grey lines in Fig.~\ref{rotationstep} show the values of $\Omega_\mathrm{c}^+$ for $n = 0,1,2,...$, for $\beta_L = 1$ (vertically offset) and $\beta_L =2$ (lower), both with $\Omega_0/2\pi = 1.26$ Hz. These values of $\beta$ were chosen to match the observed phase slips, and are consistent with calculated estimates of $L_H$ and $L_J$.  As seen, decreasing $U_\mathrm{b}$ increases $I_\mathrm{c}$ and $\beta_L$, causing $\Omega_\mathrm{c}$ to increase.

While the data show a clear signature of a phase slip from $n=0$ to $n=1$, the response of the condensate is different at higher $\Omega$. We attribute this to the fact that we have a wide ($w\approx R$) annulus.  At higher $\Omega$, the velocity mismatch between the irrotational flow ($v\propto1/r$) at the inner and outer edges and the rotating frame ($v\propto r$) makes it energetically favorable for annular vortices to appear~\cite{DonnellyStabilityPRL66, FetterLow-LyingPR67}. This is consistent with what we observe for $\Omega/2\pi>2$ Hz, as shown in Fig.~\ref{histogram}(e,f). The presence of annular vortices allows the winding number to depend on the path $P$. In particular, the difference in winding numbers at the inner and outer edges of the annulus equals the number of vortices in the annulus. This effect is reflected in the difference between the open and closed symbols of Fig.~\ref{rotationstep}. The appearance of annular vortices is analogous to the penetration of flux lines into a type-II superconductor~\cite{AndersonHardRMP64,FetterLow-LyingPR67}. The non-negligible width of the annulus also necessitates a correction to the effective radius of the ring, and therefore to $\Omega_0$. For a parabolic radial density profile with the dimensions of our trap, the correction factor is $1.06$~\cite{supplemental2}. This gives an effective $\Omega_0/2\pi = 1.26$ Hz.

In conclusion, we have deterministically driven single phase slips between circulation states in an atom circuit with a rotating weak link. This circuit is an atomic analog of a single-junction DC SQUID. The behavior of the circuit is consistent with a model assuming a $2\pi$ periodic current-phase relation for the weak link. Furthermore, by varying the barrier height, we have shown that the junction critical current can be controlled experimentally.

Other phenomena characteristic of superconducting circuits should also be observable in atomic circuits. Since rotation in our system is analogous to magnetic field in a SQUID, our device represents a proof-of principle BEC rotation sensor. Additional interesting examples include hysteresis, sensitivity to thermal effects, Shapiro steps, and multi-junction devices. Our ability to dynamically change the current-phase relation may even lead to new functionality in controlling atom circuits.

The authors thank L. Mathey, A. Mathey, C. Clark, and M. Edwards for insightful discussions, J. G. Lee and Y. Shyur for technical assistance, and S. E. Shafranjuk and M. Blamire for providing valuable information on three-terminal superconducting devices. This work was partially supported by ONR, the ARO atomtronics MURI, and the NSF PFC at JQI. CJL acknowledges support from the NIST-ARRA program.


\begin{thebibliography}{48}
\bibitem[{end({\natexlab{a}})}]{endnote:uncertainty}
\bibinfo{note}{Uncertainties herein are the uncorrelated combination of
  1$\sigma$ statistical and systematic uncertainties unless stated otherwise.}
\bibitem[{end({\natexlab{b}})}]{endnote:spontaneous}
\bibinfo{note}{The measured probability of spontaneous vortex formation was
  consistent with zero ($p<$ 5\%, with a 95\% confidence interval).}
\bibitem[{end({\natexlab{e}})}]{endnote:hysteresis}
\bibinfo{note}{For $\beta_L < \pi/2$ there are values of $\Omega$ for which
  there are no stable solutions, and dissipation can occur without leading to a
  well-defined phase slip. For $\beta_L>\pi/2$ there is a stable solution for
  all $\Omega$. In general, hysteretic behavior can occur for $\beta_L > 1$.}
\end{thebibliography}
\end{document}